\documentstyle[aps,prl,epsf,twocolumn,epsfig,floats]{revtex}

\begin{document}
\draft

%%%%%%%%%%%%%%%%%%%%%%%%%%%%%%%%%%%%%%%%%%%%%%%%%%%%%%%%%
%                                                         %
%   Comment out \wideabs{ when not using twocolumn mode   %
%                                                         %
%%%%%%%%%%%%%%%%%%%%%%%%%%%%%%%%%%%%%%%%%%%%%%%%%%%%%%%%%%

\wideabs{

\title{Doping dependence of an $n$-type cuprate superconductor investigated
by ARPES}
\author{N.P. Armitage, F. Ronning, D.H. Lu, C. Kim, A. Damascelli,
K.M. Shen, D.L. Feng, H. Eisaki, and Z.-X. Shen}
\address{Dept. of Physics, Applied Physics and Stanford Synchrotron
        Radiation Laboratory, Stanford University, Stanford, CA 94305}
\author{P.K. Mang, N. Kaneko, and M. Greven}
\address{Dept. of Applied Physics and Stanford Synchrotron
        Radiation Laboratory, Stanford University, Stanford, CA 94305}
\author{Y. Onose, Y. Taguchi, and Y. Tokura}
\address{Department of Applied Physics, The
 University of Tokyo, Tokyo 113-8656, Japan}

\date{November 20, 2001}
\maketitle
\begin{abstract}

We present an angle resolved photoemission (ARPES) doping
dependence study of the $n$-type cuprate superconductor
Nd$_{2-x}$Ce$_{x}$CuO$_{4\pm \delta}$, from the half-filled
Mott-insulator to the $T_{c}$=24K superconductor.  In
Nd$_{2}$CuO$_{4}$, we reveal the charge-transfer band (CTB) for
the first time.  As electrons are doped into the system, this
feature's intensity decreases with the concomitant formation of
near-$E_{F}$ spectral weight.  At low doping, the Fermi surface is
an electron-pocket (with volume $\sim x$) centered at ($\pi$,0).
Further doping leads to the creation of a new hole-like Fermi
surface (volume $\sim 1+x$) centered at ($\pi$,$\pi$).  These
findings shed light on the Mott gap, its doping evolution, as well
as the anomalous transport properties of the $n$-type cuprates.

\end{abstract}
} % end of \wideabs{s

The parent compounds of the high-temperature superconductors are
believed to belong to a class of materials known as Mott
insulators\cite{Anderson}. At half-filling these materials,
predicted to be metallic by band theory, are insulating due to the
large Coulomb repulsion that inhibits double site occupation and
hence charge conduction.  These cuprates become metals and then
superconductors when doped with charge carriers away from
half-filling.  Although the general systematics of Mott insulators
and metals are understood, the question of how one may proceed
from a half-filled Mott insulator (with only spin low-energy
degrees of freedom) to a metal (with a large Luttinger's theorem
respecting Fermi surface) is unanswered.  Even after 15 years of
intensive research into this fundamental issue in the cuprates,
both the manner in which this evolution occurs and the nature of
electronic states at the chemical potential remains unclear.

Within the Hubbard model, Mott insulators are described as a
single metallic band split into upper (UHB) and lower Hubbard
bands (LHB) by a correlation energy $U$ that represents the energy
cost for a site to be doubly occupied.  The high-$T_{c}$ parent
compounds are not Mott insulators $per$ $se$, but may be more
properly characterized as charge-transfer insulators\cite{Zaanen}.
However, it is believed that such systems can be described by an
effective Hubbard Hamiltonian, where an oxygen-derived charge-
transfer band (CTB) substitutes for the LHB and the charge-
transfer gap $\Delta$ plays the role of $U$.  ARPES measurements
on the prototypical half-filled parent Mott insulators
Sr$_2$CuO$_2$Cl$_2$ (SCOC) and Ca$_2$CuO$_2$Cl$_2$ (CCOC) have
shed light on the dispersive behavior of this
CTB\cite{Wells,Ronning}.  Interestingly, previous photoemission
measurements on undoped Nd$_{2}$CuO$_{4}$ (NCO) did not reveal a
similar feature\cite{Gunnarsson}.  This is surprising as one might
expect that the half-filled CuO$_2$ planes would exhibit
generalities independent of material class.  The CTB in NCO may
not have been resolved previously due to an anomalous (and yet to
be understood) polarization dependence of its intensity, and
because it is partially obscured by the main valence
band\cite{NCO}.

In the simplest picture, the chemical potential moves into the LHB
or UHB (with a transfer of spectral weight across the correlation
gap $\sim U$\cite{Meinders}), as the material is doped away from
half-filling with holes or electrons respectively.  In alternative
scenarios, the act of doping creates ``states'' inside the
insulator's gap, and as a result the chemical potential remains
relatively fixed in the middle of the gap\cite{Allen,Ino}.  A
definitive resolution of this issue has been hampered by the lack
of reliable inverse photoemission experiments that, coupled with
photoemission, could show where the $E_{F}$ states reside with
respect to the CTB and UHB.  Information obtained from
photoemission alone on different material classes has been
interpreted in terms of both scenarios\cite{Allen,Ino,Veenendal}.

The vast majority of ARPES experiments on the high-temperature
superconductors have been performed on hole-doped
materials\cite{ShenDessau,AndreaRMP}.  In contrast, the
electron-doped materials have been relatively unexplored with this
technique\cite{Allen,King,estruct,NCCOGap}. In addition to having
interesting properties of their own, these materials provide a
unique opportunity to explore the evolution from a Mott insulator
to a metal since a greater portion of the states should be
occupied, making a full view of the Mott gap possible for
photoemission experiments.

In this Letter, we report results of an ARPES study of the
electron-doped cuprate superconductor Nd$_{2-x}$Ce$_{x}$CuO$_{4\pm
\delta}$ at concentrations $x=0, 0.04, 0.10,$ and $0.15$.  For the
first time, we are able to isolate and resolve the contribution to
the spectra from the CTB on the $x=0$ sample, as has been observed
in the parent compounds of the hole-doped materials, thereby
demonstrating the universality of the electronic structure of the
CuO$_2$ plane.  In NCO this feature appears $\sim$1.3 eV below the
Fermi energy, rendering almost the entire Mott gap visible in our
experiment.  Upon doping, spectral weight is shifted from the CTB
to form states near $E_{F}$.  At very low doping, these states are
centered around ($\pi$,0), forming a small Fermi surface with
volume $\sim x$.  Simultaneously, there is an appearance of
spectral weight that begins to span and fill the insulator's gap.
At high doping levels, this in-gap spectral weight moves to
$E_{F}$ near ($\pi/2$,$\pi/2$) and thereby connects with ($\pi$,0)
derived states to form an LDA-like Fermi surface with volume $\sim
1+x$.  This evolution provides a natural explanation for the
confusing transport data from electron-doped cuprates and is
qualitatively similar to what one expects from some $t-t'-t''-U$
models.

Single crystals of Nd$_{2-x}$Ce$_{x}$CuO$_{4\pm \delta}$ (NCCO)
with concentrations $x=0, 0.04, 0.10,$ and $0.15$ were grown by
the traveling- solvent floating-zone method in 4 atm.  of O$_2$ at
Stanford University ($x=0.04, 0.10,$ and $0.15$) and The
University of Tokyo ($x=0$ and $0.15$).  Samples of $x=0$, $0.10$
and $0.15$ were oxygen reduced.  In addition some data were taken
on reduced $x=0.04$ samples and were found to have features
intermediate to unreduced $x=0.04$ and reduced $x=0.10$.
Measurements were performed at the Stanford Synchrotron Radiation
Laboratory's Beamline 5-4.  The data reported here were collected
with 16.5 eV photons with 10-20 meV energy resolution and an
angular resolution as good as 0.25$^{\circ}$ ($\sim1\%$ of the
Brillouin zone) except where indicated.  The chamber pressure was
lower than 4 x 10$^{-11}$ torr.  In all measurements, sample
temperatures were uniformly 10-20K.  This presented no charging
problems even for the $x=0$ insulator.  Cleaving the samples {\it
in situ} at 10K resulted in shiny flat surfaces, which LEED
revealed to be clean and well ordered with the same symmetry as
the bulk\cite{estruct,NCCOGap}.  No signs of surface aging were
seen for the duration of the experiments ($\sim$24 h).

In Figs.  1(a) and 1(b), we show spectra for undoped NCO with
polarization of the incoming photons at 45$^{\circ}$ to the Cu-O
bonds. Along the $\Gamma$ to ($\pi$,$\pi$) cut, a large broad
feature disperses out from under the main valence band, reaches a
maximum near ($\pi$/2,$\pi$/2), and then disperses back to higher
energy.  A very similar dispersion can be seen in the
perpendicular direction (Fig.  1(b)). A feature with a strikingly
similar shape and energy scale has been seen in CCOC, as shown in
Figs.  1(c) and 1(d).  Due to this similarity, as well the fact
that the feature's position relative to the chemical potential
($\sim$1.3 eV; minimum binding energy of its centroid) is
approximately the same as this material's optical gap $\sim$1.6 eV
(measured from the peak in the optical
conductivity)\cite{NCOoptgap}, we assign this feature to the CTB.
This assertion is supported below by its dramatic doping
dependence. There is some variability ($\pm 0.1$eV) in the exact
binding energy of the CTB from cleave to cleave, as also observed
in CCOC and SCOC.  In Figs. 1(a) and 1(b), the horizonal scale is
set to the average position from a number of cleaves.  In Figs.
1(c) and 1(d), the energy is given relative to the binding energy
minimum of the feature.  The $x=0$ sample is probably slightly
doped so that the chemical potential is pinned near the conduction
band minimum (UHB).  Part of the discrepancy between photoemission
($\sim$1.3 eV) and optics ($\sim$1.6 eV) may stem from the fact
that the charge-transfer (CT) gap appears to be indirect, as we
will show later.

\begin{figure}[htb]
\centerline{\epsfig{figure=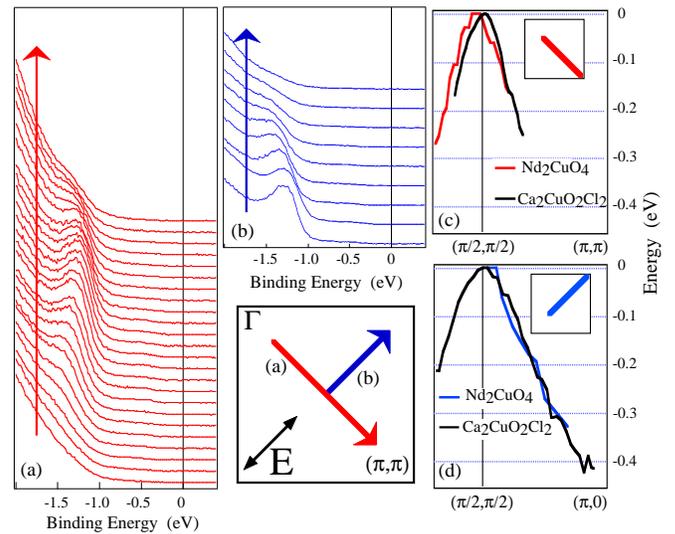,width=8.5cm}}
\vspace{.2cm} \caption{(color) (a) Dispersion of charge-transfer
(CT) band in Nd$_{2}$CuO$_{4}$ along zone diagonal from 25$\%$ to
75$\%$ of $\Gamma$ to ($\pi$,$\pi$) distance.  (b) Dispersion of
CTB from ($\pi$/2,$\pi$/2) to 50$\%$ of the ($\pi$/2,$\pi$/2) to
($\pi$,0) distance (c) and (d) Comparison of the CTB dispersion in
NCO (red and blue) and CCOC (black)}
\end{figure}

\begin{figure*}[htb]
\centerline{\epsfig{figure=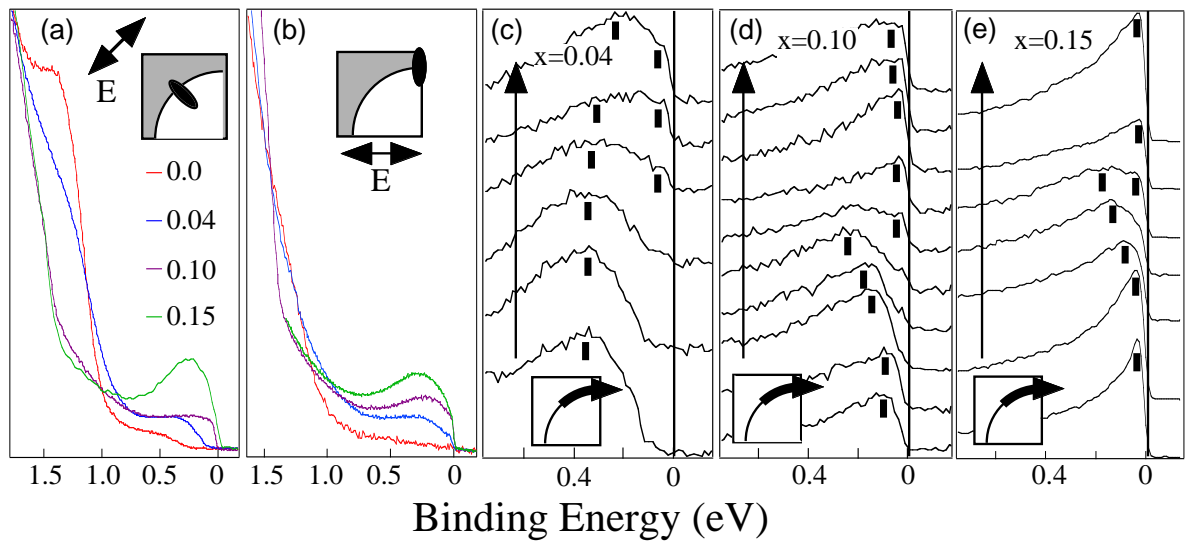,width=16cm}}
\vspace{.2cm} \caption{(a) EDCs integrated in a region near the
($\pi$/2,$\pi$/2) position. (b) EDCs integrated in a region near
the ($\pi$,0.3$\pi$) position (c), (d), and (e) EDCs from around
the putative LDA Fermi surface for $x=0.04, 0.10,$ and $0.15$
samples, respectively}
\end{figure*}

Having identified the CTB in Nd$_{2}$CuO$_{4}$, we can track it as
the dopant concentration is increased and spectral weight appears
in the near-$E_{F}$ region.  In Figs.  2(a) and 2(b), we plot
partially angle-integrated energy distribution curves (EDCs) from
regions near ($\pi$/2,$\pi$/2) and ($\pi$,$0.3\pi$) respectively.
One can see that at finite doping levels the CTB spectral weight
decreases and intensity develops at energies near $E_{F}$, at what
is ostensibly the edge of the upper Hubbard band.  This is
suggestive of the kind of transfer of spectral weight from high
energies to low energies that one qualitatively expects when
doping a Mott insulator\cite{Meinders}, and gives us confidence in
our assignment of this feature as the CTB.

A closer look at Fig.  2(a) reveals that there is a very weak
low-energy ``foot'' in the $x=0$ sample that probably reflects a
small intrinsic doping level in this otherwise half-filled
material.  The larger nondispersive near-$E_{F}$ spectral weight
that develops along the zone diagonal for the $x=0.04$ sample is
gapped by $\sim$150 meV.  This is in contrast to near
($\pi,0.3\pi$) (Fig.  2(b)), where there is spectral weight at
$E_{F}$ for $x=0.04$ doping levels.  For $x=0.10$, the zone
diagonal spectral weight has moved closer to $E_{F}$ and even
stronger $E_{F}$ intensity has formed near ($\pi$,$0.3\pi$).  A
weak dispersion is evident along the zone diagonal for the
$x=0.10$ sample (not shown).  At $x=0.15$ there is large
near-$E_{F}$ weight and a strong dispersion throughout the zone as
detailed in our previous work\cite{estruct,NCCOGap}. The small
chemical potential shift ($\sim$150 meV) seen by XPS in NCCO by
Harima et.  al. \cite{Harima} in this doping range is consistent
with our result where it appears that band at ($\pi$,0) moves on
order of this amount when doping from $x=0$ to $x=0.15.$

Following our previous analysis of the optimally doped compound
\cite{estruct}, we construct Fermi surfaces by integrating EDCs in
a small window about $E_{F}$(-40meV, +20meV) and plotting this
quantity as a function of $\vec{k}$ (Fig.  3).  Consistent with
the above observation that spectral weight along the zone diagonal
is gapped for $x=0.04$, one can see that it is only the states at
($\pi$,0) that can contribute to low-energy properties.  Here, a
Fermi ``patch'' indicates that there is an extremely low-energy
shallow band in this part of the Brillouin zone. At $x=0.10$,
weight at $E_{F}$ with low intensity begins to appear near the
zone diagonal.  Near ($\pi$,0), the ``band'' becomes deeper and
the Fermi patch becomes a Fermi surface.  At $x=0.15$, the zone
diagonal region has become intense.  The full Fermi surface has
formed and only in the intensity-suppressed regions near
($0.65\pi, 0.3\pi$) (and its symmetry- related points) at the
intersection of the Fermi surface with the  antiferromagnetic
Brillouin zone boundary does the underlying Fermi surface retain
its anomalous properties\cite{estruct}.

\begin{figure}[htb]
\centerline{\epsfig{figure=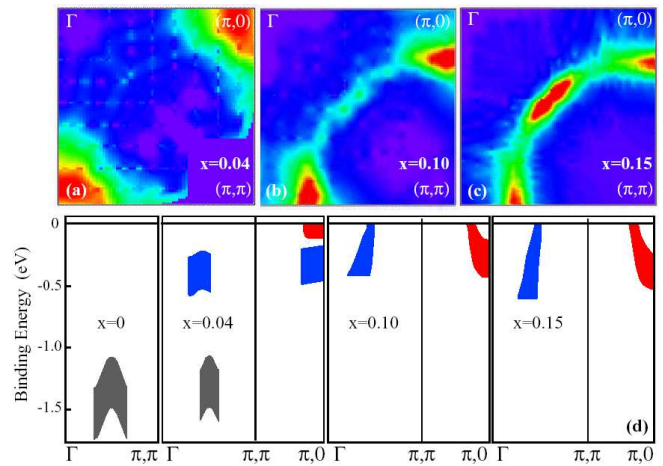,width=8.6cm}}
\vspace{.2cm} \caption{(color) Fermi surface plot: (a)
$x=0.04$,(b) $x=0.10$, and (c) $x=0.15$.  EDCs integrated in a
60meV window (-40meV,+20meV) plotted as a function of $\vec{k}$.
Data were typically taken in the displayed upper octant and
symmetrized across the zone diagonal.  (d) Panels showing doping
dependent ``band structure''. Features are plotted for the doping
levels and momentum space regions where they can be resolved.  We
do not include the slight low-energy shoulder of the $x=0$ sample,
as this is probably reflective of a small intrinsic doping level.}
\end{figure}

We can gain more insight by looking at plots of the EDCs around
the putative LDA Fermi surface, as shown in Figs.  2(c)-2(e).  In
Fig. 2(c), for $x=0.04$, a large broad feature is gapped by
$\sim$150 meV near the zone-diagonal region.  As one proceeds
around the ostensible LDA Fermi surface, the high-energy feature
loses spectral weight and may disappear, while another feature
pushes up at $E_{F}$.  It is this second component that contacts
$E_{F}$ near ($\pi$,$0.3\pi$) to form the small Fermi surface for
the $x=0.04$ sample.  Similar behavior is seen in the $x=0.10$ and
$0.15$ plots; the lowest-energy features become progressively
sharper, closer to and upon entering the metallic state.  The fact
that there are two components supports our conjecture that at low
dopings the material can be characterized by a small FS or
electron pocket around ($\pi, 0$) (volume consistent with
$x=0.04$) with doping-induced spectral weight at higher energy
elsewhere in the zone.  As the carrier concentration is increased,
the ($\pi$,0) FS deforms and a new FS segment emerges.  It derives
from the diagonal feature progressively moving to $E_{F}$, as seen
by comparing the bottom EDCs of Figs.  2(c)-2(e).  These two
segments connect to form the LDA-like Fermi surface with volume
$1.12 \pm 0.05$.

In the lower panels of Fig.  3, we present a schematic that shows
the above described evolution of the E vs.  ${\vec k}$ relation
along two symmetry directions $\Gamma-(\pi,\pi)$ and
$(\pi,\pi)-(\pi,0)$ for the four doping levels.  The centroid and
widths of the features are plotted for the doping levels and
momentum space regions where they can be resolved.

The existence of a Fermi patch at ($\pi, 0$) in the lightly doped
$x=0.04$ sample is consistent with some calculations based on the
$t-t'-t''-U$ model that predict the lowest electron-addition
states for the $x=0$ insulator are near ($\pi,
0$)\cite{numer_edoped}.  This confirms an electron-hole asymmetry
(broken by the higher order hopping terms)
\cite{numer_edoped,Kim}, as the lowest hole-addition states for
the insulator are near ($\pi/2, \pi/2$)\cite{Wells,Ronning,Kim}
and is the first direct evidence for an indirect CT gap in the
cuprates.  However, the spectral weight that appears in the the CT
gap is not explained within a simple LHB/UHB picture.  Exact
$t-t'-t''-U$ model numerical calculations have shown evidence for
intrinsic excitations to lie in the insulator's gap at low
doping\cite{numer_edoped}. These calculations show the features to
have only low spectral weight with the majority contribution at
($\pi, 0$), as is observed.  In addition, the detailed evolution
of the electronic structure with doping, especially the ${\vec k}$
space mapping of the low lying excitations of Fig.  3, bears clear
resemblance to models that allow the effective $U$ to decrease
with doping\cite{Bansil}. Within these models, the CT gap closes
with increasing doping and the Fermi level now cuts not only the
bottom of the conduction band near ($\pi, 0$), but also the top of
the valence band near ($\pi/2, \pi/2$) because of the indirect
gap\cite{Bansil}.  Here, the band-structure changes as the
antiferromangetic coherence factors and gap subside.

Starting from the metallic side, an alternative approach may be
one that emphasizes a coupling of electrons to magnetic (or
similar) fluctuations with characteristic wavevector ($\pi,\pi$).
For example, within this picture as the antiferromagnetic phase is
approached and antiferromagnetic correlations become stronger the
``hot spot'' regions (intersection of the FS and antiferromagnetic
Brillouin zone boundary \cite{estruct}) may spread so that the
zone-diagonal spectral weight is gapped by the approximate nesting
of the ($\pi/2, \pi/2$) section of FS with the ($-\pi/2,-\pi/2$)
section of FS.  This scheme obviously breaks down as one gets
close to the Mott state, where the zone diagonal spectral weight
is not only gapped, but also vanishes.

Our finding of an electron pocket that evolves with doping into a
large hole-like Fermi surface provides a route towards explaining
the long-standing puzzle that while transport in these materials
exhibit unambiguous $n$-type carrier behavior at low doping, one
has to invoke both electron and hole-carriers to explain data near
optimal doping \cite{Wang}.

In conclusion, it appears that certain elements of both scenarios
laid out in the introduction can explain our data.  At low carrier
concentrations, electrons are doped into regions close to
($\pi,0$) (confirming a particle-hole asymmetry) near the energy
expected for the UHB, forming a small Fermi surface.
Simultaneously, there is an appearance of spectral weight at
higher energy that begins to span and fill the insulator's gap. At
higher electron doping levels, more spectral weight is created in
this midgap region and it is this high-energy spectral weight that
moves to the chemical potential and completes the ${\vec k_F}$
segment to form a large Fermi surface with a volume close to the
expected Luttinger volume.

The authors would like to thank T.  Tohyama, P.G.  Steeneken, C.
Kusko, and R.S.  Markiewicz for valuable correspondences.
Experimental data was recorded at SSRL which is operated by the
DOE Office of Basic Energy Science, Div.  of Chem.  Sciences and
Mat.  Sciences.  Additional support comes from the Office of Naval
Research.  The crystal growth work at Tokyo was supported in part
by Grant-in-Aids for Scientific Research from the Ministry of
Education, Science, Sports, and Culture, Japan, and NEDO.  The
Stanford crystal growth was supported by the U.S.  Department of
Energy under contracts No.  DE-FG03-99ER45773 and No.
DE-AC03-76SF00515, by NSF CAREER Award No.  DMR-9985067, and by
the A.P.  Sloan Foundation.

\end{document}